\documentclass[twocolumn]{aastex631}
\usepackage{xspace} 
\usepackage{graphicx} 
\usepackage{natbib}
\usepackage{amsmath}
\newcommand{\y}[1]{\ensuremath{y_{#1}}\xspace}

\newcommand{\gl}{\ell}
\newcommand{\gb}{{\it b}}

\newcommand{\kms}{\ensuremath{\,{\rm km\,s^{-1}}}\xspace}

\newcommand{\m}{\ensuremath{\,{\rm m}}\xspace}

\newcommand{\K}{\ensuremath{\,{\rm K}}\xspace}

\newcommand{\h}[1]{\ensuremath{^{#1}{\rm H}}\xspace}
\newcommand{\hi}{H\,{\sc i}}

\newcommand{\vect}[1]{\boldsymbol{#1}}
\submitjournal{The Astrophysical Journal}
\received{March 14, 2024}
\revised{March 27, 2024}
\accepted{March 27, 2024}

\shorttitle{\hi\ Absorption Scale Height}
\shortauthors{Wenger et al.}

\begin{document}

\title{Revisiting the Vertical Distribution of \hi\ Absorbing Clouds in the
  Solar Neighborhood}

\author[0000-0003-0640-7787]{Trey V. Wenger} 
\affiliation{NSF Astronomy \& Astrophysics Postdoctoral Fellow,
  Department of Astronomy, University of Wisconsin--Madison,
  Madison, WI, 53706, USA}
\email{twenger2@wisc.edu}

\author[0000-0003-3351-6831]{Daniel R. Rybarczyk} 
\affiliation{NSF Astronomy \& Astrophysics Postdoctoral Fellow,
  Department of Astronomy, University of Wisconsin--Madison,
  Madison, WI, 53706, USA}

\author[0000-0002-3418-7817]{Sne\v{z}ana Stanimirovi\'{c}}
\affiliation{Department of Astronomy, University of
  Wisconsin--Madison, Madison, WI, 53706, USA}

\begin{abstract}
  The vertical distribution of cold neutral hydrogen (\hi) clouds is a
  constraint on models of the structure, dynamics, and hydrostatic
  balance of the interstellar medium. In 1978, Crovisier pioneered a
  method to infer the vertical distribution of \hi\ absorbing clouds
  in the solar neighborhood. Using data from the Nan\c{c}ay 21-cm
  absorption survey, they determine the mean vertical displacement of
  cold \hi\ clouds, \(\langle|z|\rangle\). We revisit Crovisier's
  analysis and explore the consequences of truncating the
  \hi\ absorption sample in Galactic latitude. For any non-zero
  latitude limit, we find that the quantity inferred by Crovisier is
  not the mean vertical displacement but rather a ratio involving
  higher moments of the vertical distribution. The resultant
  distribution scale heights are thus \({\sim}1.5\) to \({\sim}3\)
  times smaller than previously determined. In light of this
  discovery, we develop a Bayesian Monte Carlo Markov Chain method to
  infer the vertical distribution of \hi\ absorbing clouds. We fit our
  model to the original Nan\c{c}ay data and find a vertical
  distribution moment ratio \(\langle|z|^3\rangle/\langle|z|^2\rangle
  = 97 \pm 15\,\text{pc}\), which corresponds to a Gaussian scale
  height \(\sigma_z = 61 \pm 9\,\text{pc}\), an exponential scale
  height \(\lambda_z = 32 \pm 5\,\text{pc}\), and a rectangular
  half-width \(W_{z, 1/2} = 129 \pm 20\,\text{pc}\). Consistent with
  recent simulations, the vertical scale height of cold \hi\ clouds
  appears to remain constant between the inner-Galaxy and the
  Galactocentric distance of the solar neighborhood. Local
  fluctuations might explain the large scale height observed at the
  same Galactocentric distance on the far side of the Galaxy.
\end{abstract}

\section{Introduction}

The structure of the interstellar medium (ISM) is determined by the
complex interplay of many physical processes
\citep[e.g.,][]{mckee1977}. In particular, the vertical distribution
(i.e., perpendicular to the Galactic plane) of interstellar matter is
set by the local gravitational potential, energy injection by stellar
feedback and supernovae, the radiation field, and the physical
conditions (e.g., pressure, turbulence, magnetic support) of the
ISM \citep[e.g.,][]{kim2010,ostriker2022}. Characterizing the vertical
distribution of the various phases of the ISM ultimately constrains
models and simulations of ISM physics, star formation, and galactic
evolution \citep[e.g.,][]{hopkins2018}.

Neutral atomic hydrogen (\hi) is pervasive in the ISM and exists in
pressure equilibrium in two stable phases: the cold neutral medium
(CNM) and the warm neutral medium (WNM)
\citep{field1969,mckee1977,wolfire2003}. It is within clumps of CNM
gas that molecular clouds, and ultimately stars, form
\citep{mckee2007}. The WNM is easily traced via the 21 cm hyperfine
transition of \hi\ in emission, and its vertical distribution has been
extensively studied. \citet{levine2006}, for example, find a strong
anti-correlation between the vertical scale height of \hi\ emission
and spiral structure inferred from the \hi\ surface density. This
suggests that the gravitational potential of spiral arms acts to
squeeze the WNM \citep[e.g.,][]{kalberla2009}. Given its low kinetic
and spin temperature (\(T \sim 100\,\text{K}\)), the CNM is most
easily observed via 21 cm absorption toward background continuum
sources \citep[e.g.,][]{dickey1983} or in self-absorption toward
background \hi\ emission \citep[e.g.,][]{kavars2003}. Such
\hi\ absorption observations are challenging to interpret in the Milky
Way ISM due to the blending of \hi\ features in both emission and
absorption along sight-lines through the Galactic midplane
\citep[e.g.,][and references
  therein]{dickey2003,mcclure-griffiths2023}. Nonetheless, several
studies have explored the differences between the vertical
\hi\ distribution as traced by \hi\ emission and
absorption. \citet{dickey2009}, for example, explore the
Galactocentric radial and azimuthal variation in \hi\ scale height in
the outer Galaxy. They find similar scale heights between the two
\hi\ phases.

In the inner Galaxy, \citet{dickey2022} use \hi\ absorption
observations from the Galactic Australia Square Kilometer Array
Pathfinder (GASKAP) pilot survey to constrain the vertical scale
height of CNM clouds. The complications of interpreting
\hi\ absorption spectra in the inner Galaxy are mitigated by
considering only those components near the tangent point, where
distance ambiguities are minimized. They find a compact vertical CNM
cloud distribution, with an inferred scale height \(\sigma_z \simeq
50-90\,\text{pc}\). This distribution closely matches the scale height
of the thin molecular cloud layer in the inner Galaxy, \(\sigma_z
\simeq 40-60\,\text{pc}\)
\citep{bronfman1988,heyer2015,su2021}. Furthermore, \citet{dickey2022}
infer a scale height of \(\sigma_z \simeq 200-700\,\text{pc}\) for
\hi\ absorbing clouds beyond the solar circle in the fourth Galactic
quadrant, which suggests significant flaring of the CNM in the
Galactic outskirts. This is consistent with Galactic simulations where
the \hi\ disk becomes more vertically extended at the edges of
galaxies due to the decreased pressure of the ISM, although the effect
is seen only in the warm \hi\ and not the CNM \citep{smith2023}.

\citet{dickey2022} also estimate the vertical scale height of CNM
clouds near the solar orbit by considering those \hi\ absorption
detections with velocities relative to the local standard of rest
(LSR) near \(0\kms\). At these velocities, their sample includes both
local \hi\ absorption as well as \hi\ absorption near the solar
Galactocentric distance, \(R_0\), along a Galactic longitude \(\ell
\simeq 340^\circ\), which, for \(R_0 = 8.34\,\text{kpc}\)
\citep{reid2014}, corresponds to a heliocentric distance \(d \simeq
15.6\,\text{kpc}\). They find that the latitude distribution of these
\hi\ absorption features is a blend of two Gaussian distributions with
different widths, presumably corresponding to the local and distant
CNM clouds. Considering only the narrow component, they estimate a
vertical scale height of \(\sigma_z \simeq 160\,\text{pc}\) for the
CNM at the solar orbit.

There are few experiments investigating the vertical distribution of
the CNM in the solar neighborhood. In fact, nearly every reference
that we could find in the literature ultimately pointed back to a
single experiment: \citet{crovisier1978}. In this work, the authors
develop a statistical model to infer the vertical scale height of
local \hi\ absorbing clouds. Using data from the Nan\c{c}ay 21-cm
absorption survey \citep{nancay1978}, \citet{crovisier1978} constrains
the mean absolute vertical displacement of CNM clouds in the solar
neighborhood as \(\langle|z|\rangle = 107 \pm 29\,\text{pc}\). They
claim that the data are inconsistent with an exponential vertical
distribution, and the inferred \(\langle|z|\rangle\) implies a scale
height \(\sigma_z = 134\pm36\,\text{pc}\) for a Gaussian vertical
distribution. \citet{belfort1984} expand this experiment to include
more \hi\ absorption data, but the estimated mean absolute vertical
displacement is similar: \(\langle|z|\rangle = 92\pm12\,\text{pc}\)
for clouds with \(5^\circ < |b| < 30^\circ\). These results are
consistent with the \citet{dickey2022} inference at \(R_0\) in the
fourth Galactic quadrant. Altogether we begin to develop a clear
picture of a vertical CNM distribution that is thin in the
inner-Galaxy, mimicking the molecular gas distribution, and slowly
flares with Galactocentric distance \citep[e.g., see Section
  6.4.2. and Figure 10 in][]{mcclure-griffiths2023}.

In this work, we revisit the \citet{crovisier1978} analysis in
preparation for its application to the ever-growing sample of
\hi\ absorption spectra \citep[e.g., {\sc
    hibigcat},][]{mcclure-griffiths2023} and next-generation
\hi\ surveys \citep[e.g., GASKAP,][]{dickey2013}. In particular, we
discover a statistical bias that alters the interpretation of the
\citet{crovisier1978} result, which we demonstrate through both a
simple simulation as well as a robust derivation. We develop a
Bayesian model that performs a similar inference as in the original
work and, using the original Nan\c{c}ay data, implement a Monte Carlo
Markov Chain (MCMC) method to infer the vertical distribution of CNM
clouds in the solar neighborhood. Finally, we discuss the implications
of these results in light of modern Galactic ISM simulations.

\section{Method}\label{sec:crovisier}

Our goal is to infer the shape of the vertical distribution of
\hi\ absorbing clouds (or any other tracer) using only their sky
positions and radial velocities. \citet{crovisier1978} introduce a
statistical method to perform this inference, which we briefly
summarize here (see their Section 2 for more details). For convenience
and simplicity, we adopt a slightly different notation: \(d\) is the
heliocentric distance and \(z \equiv |z|\) is the absolute vertical
displacement of the cloud from the Galactic midplane. We assume that
the Sun is located at \(z = 0\) and the midplane is defined by
Galactic latitude \(\gb = 0^\circ\).

The observed radial velocity of a cloud is related to its distance due
to Galactic rotation. For a cloud located at a given Galactic
longitude and latitude, (\(\gl, \gb\)), the radial velocity in the LSR
frame is
\begin{equation}
  V_{\rm LSR} = V_{\odot, \rm LSR}(\gl, \gb) + V_{\rm rot}(d, \gl, \gb) + V_t
\end{equation}
where \(V_{\odot, \rm LSR}\) is the solar motion with respect to the
LSR projected onto the line of sight, \(V_{\rm rot}\) is the cloud's
motion due to Galactic rotation projected onto the line of sight, and
\(V_t\) is the random velocity of the cloud along the line of
sight. \citet{crovisier1978} use a local approximation for Galactic
rotation defined by the first Oort constant, but any model that
predicts \(V_{\rm rot}\) for a given distance and sky position is
appropriate.

Given that radial velocity is only a weak indicator of distance in the
solar neighborhood, \(d\) cannot be determined for any individual
cloud with sufficient precision to infer the shape of the vertical
distribution of clouds. If we assume a plane-parallel model for clouds
in the solar neighborhood, then, \citet{crovisier1978} argues, we can
relate the first moment of a cloud's distance probability distribution
(i.e., the expectation value), \(\langle d\rangle\), to the first
moment of the vertical distribution of clouds (i.e., the mean absolute
vertical displacement), \(\langle |z|\rangle\), by
\begin{equation}
  \langle d\rangle = \frac{\langle |z|\rangle}{\sin |\gb|}. \label{eq:mistake}
\end{equation}
With a sufficiently large sample of clouds, the central limit theorem
suggests that we can infer \(\langle |z|\rangle\) from these distance
point estimates despite their imprecision.

Finally, we assume that \(V_t\) is truly random; that is, we assume
that there are no systematic departures from the Galactic rotation
model in our sample of clouds. \citet{crovisier1978} uses a
least-squares algorithm to minimize
\begin{equation}
  \sum V_t^2 = \sum \left[V_{\rm LSR} - V_{\odot, \rm LSR} - V_{\rm rot}(\langle d\rangle)\right]^2, \label{eq:least_squares}
\end{equation}
where the sum is taken over all of the clouds, in order to infer the
mean absolute vertical displacement, \(\langle |z|\rangle\), the
root-mean-square of \(V_t\), and the parameters that define the solar
non-circular motion (\(V_{\odot, \rm LSR}\)) and Galactic rotation
(\(V_{\rm rot}\)).

\subsection{The Bias}\label{sec:mistake}

We must first derive equation~\ref{eq:mistake}, which relates the
first moment of a cloud's distance probability distribution to the
first moment of the vertical distribution of clouds, before we can
demonstrate the bias introduced by a subtle assumption.  Consider a
population of clouds that follows an absolute vertical displacement
distribution defined by the probability density \(P_Z(z)\). We adopt a
notation where the arguments of \(P\) represent random variables and
the subscripts of \(P\) represent the space over which the probability
is defined. In this case, \(z \in Z = \mathbb{R}^+\). Assuming that
the clouds are distributed uniformly across the Galactic plane (i.e.,
a plane-parallel model), then the distance probability density for a
cloud with Galactic latitude \(\gb\) is
\begin{equation}
  P_D(d) = P_Z(z)\frac{dz}{dd} = P_Z(z)\sin|\gb|
\end{equation}
where \(z = d\sin|\gb|\). Thus, \(P_Z(z) = P_D(d) / \sin|\gb|\) for
\(\sin|\gb| > 0\). The first moment of the distance probability
distribution is
\begin{align}
  \langle d\rangle & = \int_0^\infty dP_D(d)\,dd \nonumber \\
  & = \frac{1}{\sin|\gb|}\int_0^\infty zP_Z(z)\,dz \nonumber \\
  & = \frac{\langle |z|\rangle}{\sin|\gb|} \label{eq:derivation}
\end{align}
for \(\sin|\gb| > 0\), where \(\langle |z|\rangle\) is the first
moment of the absolute vertical displacement distribution. Presumably,
this is how \citet{crovisier1978} arrive at equation~\ref{eq:mistake}.

\begin{figure}
  \centering
  \includegraphics[width=\linewidth]{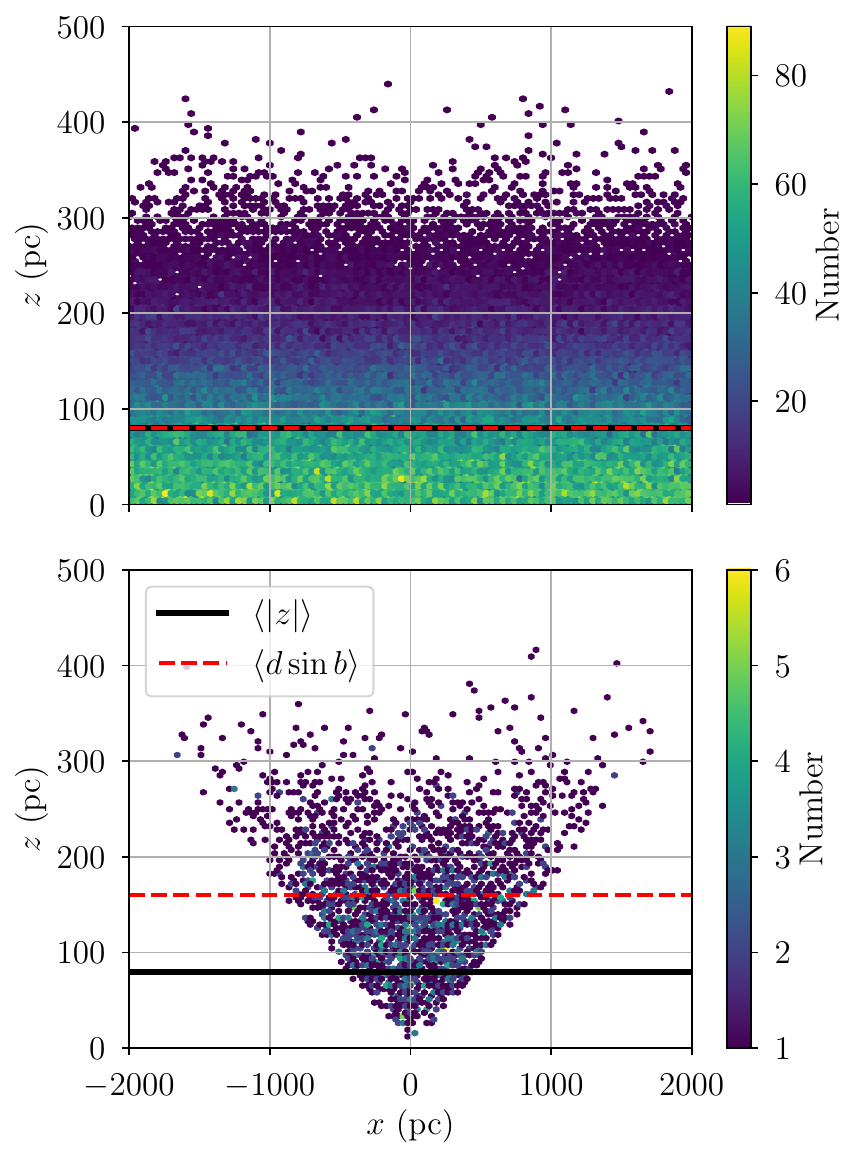}
  \caption{The binned heliocentric \((x, z)\) distribution of
    simulated clouds drawn from a uniform distribution in \(x\) and
    \(y\) and a half-normal distribution in \(z\) with standard
    deviation \(\sigma_z = 100\,\text{pc}\). Note that the scale of
    the \(x\) axis is much larger than that of the \(z\) axis. The top
    panel includes all clouds whereas the bottom panel only includes
    clouds with \(b > b_{\rm min} = 10^\circ\). The solid black lines
    indicate \(\langle|z|\rangle = \sigma_z\sqrt{2/\pi} \simeq
    80\,\text{pc}\) for a half-normal distribution. The dashed red
    lines indicate \(\langle d\sin|b|\rangle\) derived from the
    simulated clouds. Clearly, \(\langle|z|\rangle \neq \langle
    d\sin|b|\rangle\) for \(b_{\rm min} >
    0^\circ\). \label{fig:simulation}}
\end{figure}

A simple simulation reveals a problem with equation~\ref{eq:mistake}
when we truncate the data in Galactic latitude. \citet{crovisier1978}
motivates the need for latitude truncation in order to (1) restrict
the sample to local \hi\ clouds and (2) remove complex \hi\ absorption
spectra in the Galactic plane. We demonstrate the problem by drawing a
random sample of clouds from a uniform distribution in the Galactic
plane out to some maximum midplane distance, \(R_{\rm max}\), such
that \((x, y) \sim U(-R_{\rm max}, R_{\rm max})\), and a half-normal
distribution in absolute vertical displacement with some variance,
\(\sigma_z^2\), such that \(z \sim N_{1/2}(\sigma_z^2)\). For a
half-normal distribution, the standard deviation is related to the
first moment by \(\langle|z|\rangle = \sigma_z\sqrt{2/\pi}\). The
midplane distance is \(r = \sqrt{x^2 + y^2}\) and the Galactic
latitude is \(\gb = \tan^{-1}(z/r)\). For a random sample of 1,000,000
clouds with \(R_{\rm max} = 10\,\text{kpc}\) and \(\sigma_z =
100\,\text{pc}\), we use equation~\ref{eq:mistake} to calculate
\(\langle|z|\rangle = \langle d\sin|\gb|\rangle \simeq
80\,\text{pc}\), which is consistent with the expectation for a
half-normal vertical distribution, \(\langle|z|\rangle =
\sigma_z\sqrt{2/\pi} \simeq 80\,\text{pc}\). If we remove the clouds
with \(|\gb| < \gb_{\rm min}\), however, then the discrepancy
emerges. For the same simulated data with \(\gb_{\rm min} = 1^\circ\),
for example, we find \(\langle d\sin|\gb|\rangle \simeq
120\,\text{pc}\). For \(\gb_{\rm min} \geq 2^\circ\), we find
\(\langle d\sin|\gb|\rangle \simeq 160\,\text{pc}\). There is a factor
of two difference from the equation~\ref{eq:mistake} prediction that
appears to be independent of \(\gb_{\rm min}\) for \(\gb_{\rm min}
\geq 2^\circ\) (i.e., for \(R_{\rm max} \gg \sigma_z/\tan b_{\rm
  min}\)). Figure~\ref{fig:simulation} demonstrates the result of this
simple simulation.

\subsection{The Correction}

Equation~\ref{eq:mistake} is only correct for \(b_{\rm min} =
0^\circ\).  For any \(b_{\rm min} > 0^\circ\), we must derive the
distance expectation value while properly considering the joint
dependency between \(d\), \(z\), and \(\gb\). Consider again a
plane-parallel geometry where clouds are distributed uniformly in the
Galactic midplane out to a maximum midplane distance \(R_{\rm max}\)
and follow some absolute vertical displacement distribution
\(P_Z(z)\). The joint probability distribution for a cloud's location
in the heliocentric Galactic Cartesian frame is
\begin{align}
  P_{XYZ}(x, y, z) & = P_X(x)P_Y(y)P_Z(z) \nonumber \\
  & = \frac{1}{\pi R_{\rm max}^2}P_Z(z).
\end{align}
The transformation to the cylindrical frame is
\begin{equation}
  P_{RLZ}(r, \gl, z) = \frac{r}{\pi R_{\rm max}^2}P_Z(z)
\end{equation}
and to the spherical frame is
\begin{equation}
  P_{DLB}(d, \gl, \gb) = \frac{d^2\cos^3\gb}{\pi R_{\rm max}^2}P_Z(d\sin|\gb|).
\end{equation}
Marginalized over distance, the probability density is
\begin{align}
  P_{LB}(\gl, \gb) & = \int_0^{R_{\rm max}/\cos\gb}P_{DLB}(d, \gl, \gb)\,dd \nonumber \\
  & = \int_0^{R_{\rm max}\tan|\gb|}\frac{z^2\cos\gb}{\pi R_{\rm max}^2\sin^3|\gb|}P_Z(z)\,dz.
\end{align}
If we require that \(P_Z(z) \rightarrow 0\) as \(z \rightarrow R_{\rm
  max}\tan|\gb|\) (i.e., the artificially truncated midplane
distribution does not appreciably truncate the \(z\) distribution
along the line of sight toward \(\gb\)), then
\begin{equation}
  P_{LB}(\gl, \gb) \simeq \frac{\cos\gb}{\pi R_{\rm max}^2\sin^3|\gb|}\langle|z|^2\rangle
\end{equation}
where \(\langle|z|^2\rangle = \int_0^\infty z^2P_Z(z)\,dz\) is the
second moment of the \(z\) distribution. The conditional probability
distribution -- the probability of finding a cloud at a given distance
along a given line of sight -- is
\begin{align}
  P_{D|LB}(d | \gl, \gb) & = \frac{P_{DLB}(d, \gl, \gb)}{P_{LB}(\gl, \gb)} \nonumber \\
  & = \frac{d^2\sin^3|\gb|}{\langle|z|^2\rangle}P_Z(d\sin|\gb|).
\end{align}
The first moment of this distribution is
\begin{equation}
  \langle d | \gl, \gb\rangle = \int_0^{R_{\rm max}/\cos\gb}dP_{D|LB}(d | \gl, \gb)\,dd.
\end{equation}
If we again require that \(P_{D|LB}(d | \gl, \gb) \rightarrow 0\) as
\(d \rightarrow R_{\rm max}/\cos\gb\) (i.e., the artificially
truncated midplane distribution does not appreciably truncate the
distance distribution along the line of sight toward \(\gb\)), then
\begin{equation}
  \langle d | \gl, \gb\rangle \simeq \frac{\langle|z|^3\rangle}{\langle|z|^2\rangle\sin|\gb|}. \label{eq:corrected}
\end{equation}
Thus we find that, for any \(b_{\rm min} > 0^\circ\), the distance
expectation value is not directly related to the first moment of the
absolute vertical displacement distribution, \(\langle|z|\rangle\), as
in equation~\ref{eq:mistake}, but rather the ratio of the third moment
(skewness) to the second moment (variance),
\(\langle|z|^3\rangle/\langle|z|^2\rangle\).

Equipped with this corrected derivation, we can test the result
against our simple simulation. For a half-normal distribution,
\(\langle|z|\rangle = \sigma_z\sqrt{2/\pi}\), \(\langle|z|^2\rangle =
\sigma_z^2\), and \(\langle|z|^3\rangle =
2\sigma_z^3\sqrt{2/\pi}\). For our simulation parameters \(R_{\rm max}
= 10\,\text{kpc}\) and \(\sigma_z = 100\,\text{pc}\), the assumptions
of our derivation are valid for \(b_{\rm min} \geq 2^\circ\) since
\(P_Z(R_{\rm max}\tan b_{\rm min}) \simeq 0\) and \(P_{D|LB}(R_{\rm
  max}/\cos b_{\rm min} | \gl, \gb_{\rm min}) \simeq 0\). The moment
ratio is related to the standard deviation for a half-normal
distribution by
\begin{equation}
  \langle d\sin|\gb|\rangle \simeq \frac{\langle|z|^3\rangle}{\langle|z|^2\rangle} = 2\langle|z|\rangle = 2\sqrt{\frac{2}{\pi}}\sigma_z.
\end{equation}
Here we see the erroneous factor of two. For our simulation
parameters, we predict \(\langle d\sin|\gb|\rangle \simeq
160\,\text{pc}\), which is consistent with our simulation results.

For the other vertical distribution shapes considered by
\citet{crovisier1978}, we calculate the factor by which they
overestimate the distribution width due to latitude truncation, which
is simply the ratio
\((\langle|z|^3\rangle/\langle|z|^2\rangle)/\langle|z|\rangle\).  As
demonstrated for the half-normal distribution, this factor is
\(2\). For an exponential distribution with scale height
\(\lambda_z\), the moments are \(\langle|z|^n\rangle =
n!\lambda_z^n\), and thus the factor is \(3\).  Finally, for a
rectangular distribution parameterized by the half-width, \(W_{z,
  1/2}\), the moments are \(\langle|z|^n\rangle = W_{z,
  1/2}^n/(n+1)\), so the factor is \(1.5\). Thus, for a Gaussian,
exponential, or rectangular distribution, \citet{crovisier1978}
overestimates the distribution width by a factor of 2, 3, or \(1.5\),
respectively, due to the subtle bias introduced by latitude
truncation.

\startlongtable
\begin{deluxetable}{lDDD@{\(\pm\)}D}
  \tablecaption{\hi\ Absorption Sample\label{tab:data}}
  \tablehead{
    \colhead{﻿Source Name} & \multicolumn{2}{c}{\(\gl\)} & \multicolumn{2}{c}{\(\gb\)} & \multicolumn{4}{c}{\(V_{\rm LSR}\)} \\
    \colhead{} & \multicolumn{2}{c}{deg} & \multicolumn{2}{c}{deg} & \multicolumn{4}{c}{\(\kms\)}
  }
  \decimals
  \startdata
0007+12 & 107.51 & -48.89 & -8.1 & 1.1 \\
0013+79 & 121.28 & 16.54 & 1.4 & 0.6 \\
0026+34 & 117.86 & -27.71 & -3.3 & 0.4 \\
0038+32 & 120.47 & -29.65 & -1.2 & 1.2 \\
0038+09 & 118.72 & -52.73 & -7.8 & 0.2 \\
\enddata
\tablenotetext{a}{These absorption parameters were determined ``by eye'' by \citet{nancay1978}, so we assign a LSR velocity uncertainty of \(1.0\kms\).}
\tablecomments{Table 1 is published in its entirety in the machine-readable format.
      A portion is shown here for guidance regarding its form and content.}
\end{deluxetable}

\section{Data \& Analysis}

Here we use the original \citet{nancay1978} data to reanalyze the
vertical distribution of \hi\ absorpting clouds in the solar
neighborhood in light of the aforementioned latitude truncation
bias. Our objectives are to (1) reproduce the \citet{crovisier1978}
original results, (2) perform a similar least squares analysis using a
modern Galactic rotation model, and (3) infer the vertical cloud
distribution using a robust, outlier-tolerant Bayesian model that
incorporates the uncertainties in the assumed Galactic rotation model.

\subsection{Data}

\begin{figure}
  \centering
  \includegraphics[width=\linewidth]{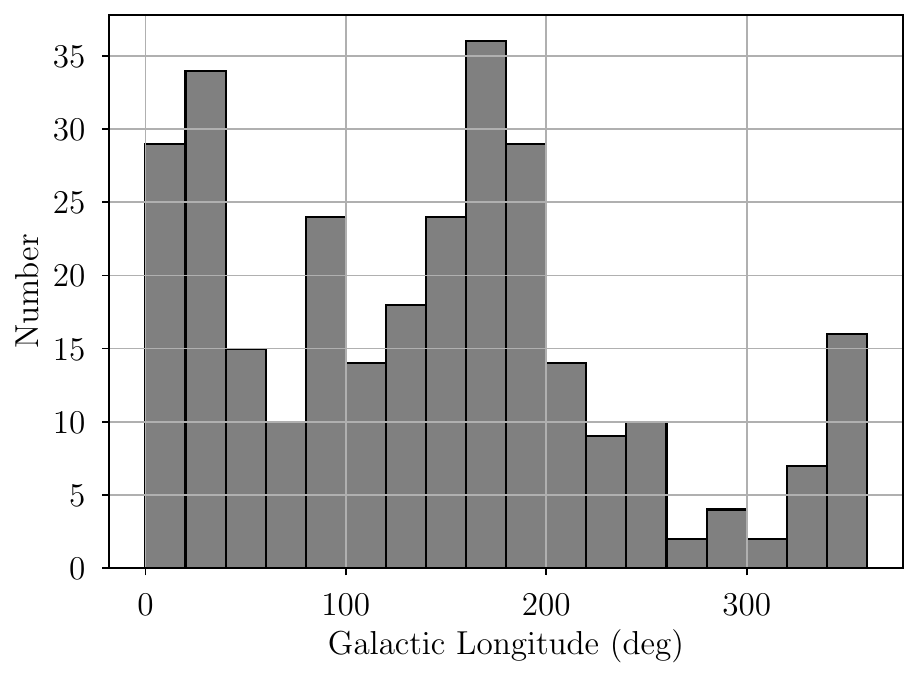}
  \includegraphics[width=\linewidth]{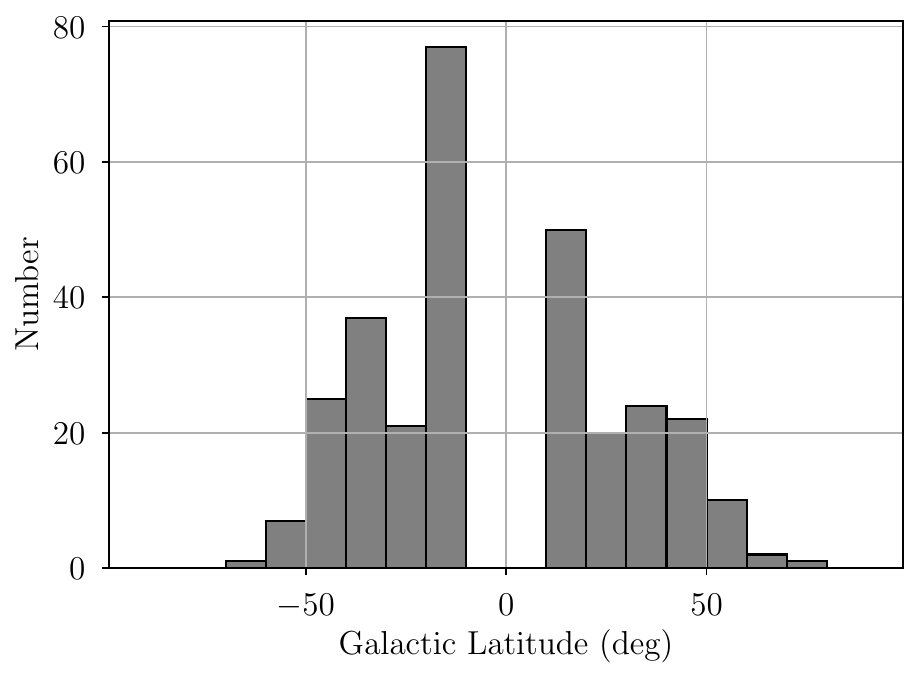}
  \includegraphics[width=\linewidth]{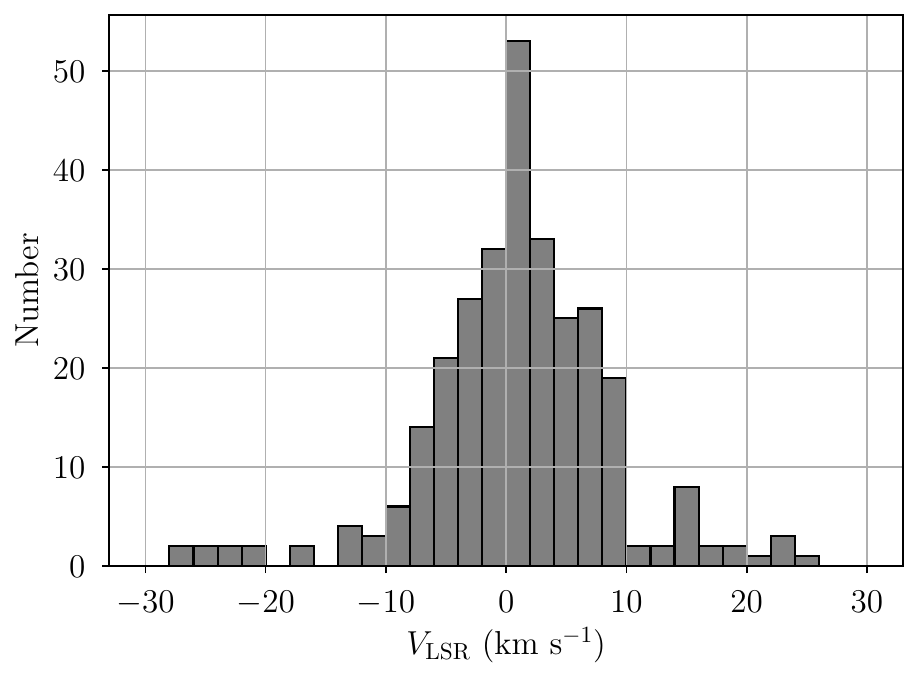}
  \caption{Galactic longitude (top), latitude (middle), and LSR
    velocity, \(V_{\rm LSR}\), (bottom) distributions of the
    \citet{nancay1978} \hi\ absorption detections. Three detections
    with \(V_{\rm LSR} < -40\kms\) are excluded from the bottom panel
    for clarity.\label{fig:crovisier}}
\end{figure}

The \citet{crovisier1978} data come from the Nan\c{c}ay 21-cm
absorption survey \citep{nancay1978}. At 21 cm, the Nan\c{c}ay radio
telescope has a half-power beam with of \(3^\prime.1\) in right
ascension and \(21^\prime\) in declination. The survey results contain
\hi\ absorption spectra toward 819 extragalactic continuum sources
with a spectral resolution of \(6\,\text{kHz}\)
(\({\sim}1.25\kms\)). Unfortunately, we were unable to find these data
in a machine readable format, so we transcribed the data by hand from
Table 2 in \citet{nancay1978}. Following the selection criteria in
\citet{crovisier1978}, we only copied rows for which the Galactic
latitude is \(|\gb| > 10^\circ\) and the absorption detection is not
marked as ``dubious'' (remark ``P'' in their Table 2). The source
names, Galactic coordinates, and LSR velocities for these 297
absorption features are listed in
Table~\ref{tab:data}. \citet{crovisier1978} claims to have 299
absorption detections meeting these criteria, but we are unable to
reproduce this number. The absorption parameters for three detections
(two toward 0538+49 and one toward 0725+14) were determined ``by eye''
by \citet{nancay1978}, and thus do not have an associated velocity
uncertainty. Figure~\ref{fig:crovisier} shows the Galactic longitude,
latitude, and LSR velocity distributions for this sample. There are
minor differences between our LSR velocity distribution and the
similar distribution in \citet{crovisier1978} (the top panel of their
Figure 1), which we attribute to the slightly different sample size of
our transcribed data.

\subsection{Least Squares}

We begin our analysis using the least squares method described by
\citet{crovisier1978} in order to reproduce their results and verify
the accuracy of the transcribed data. Following \citet{crovisier1978},
we adopt a local approximation for Galactic rotation defined in terms
of Oort's \(A\) constant,
\begin{equation}
  V_{\rm rot}(d) = Ad\sin\left[2(\gl -\gl_0)\right]\cos^2\gb,
\end{equation}
where \(\ell_0\) is the nodal deviation and \(A = 15\,\text{km
  s\(^{-1}\) kpc\(^{-1}\)}\). The solar motion with respect to the LSR
is defined as
\begin{equation}
  V_{\odot,\rm LSR} = \left(U_\odot\cos\gl + V_\odot\sin\gl\right)\cos\gb + W_\odot\sin\gb
\end{equation}
where \(U_\odot\), \(V_\odot\), and \(W_\odot\) are the solar motion
velocities in the direction of the Galactic center, solar orbit, and
north Galactic pole, respectively. We assign each cloud to its
distance expectation value (equation~\ref{eq:corrected}) given the
\(z\) distribution moment ratio,
\(\langle|z|^3\rangle/\langle|z|^2\rangle\), and we use a least
squares algorithm to search for the solar motion velocities, \(z\)
distribution moment ratio, and nodal deviation that minimize the
velocity residuals (equation~\ref{eq:least_squares}). Note that,
following \citet{crovisier1978}, we ignore the LSR velocity
uncertainties of individual \hi\ absorption detections since the
velocity residuals are dominated by the random component, \(V_t\). The
results of this analysis, applied to the same Galactic
latitude-selected and LSR velocity-selected subsets of data listed in
Tables 1 and 2 of \citet{crovisier1978}, are presented in
Table~\ref{tab:least_squares}. For each subset, we list the number of
absorption features, \(N\), optimized model parameters, \(U_\odot\),
\(V_\odot\), \(W_\odot\), \(\gl_0\), and
\(\langle|z|^3\rangle/\langle|z|^2\rangle\), and root mean square
residual velocity, \(\sigma_V\).

As expected, the results of our reproduction of the
\citet{crovisier1978} least squares analysis are nearly identical to
the original results, except that we now correctly interpret the
inferred quantity as \(\langle|z|^3\rangle/\langle|z|^2\rangle\)
rather than \(\langle|z|\rangle\). This similarity suggests that we
have accurately transcribed the original data and have implemented the
least squares analysis correctly. We are unable to rectify the minor
differences in sample size, and we attribute the small differences in
inferred values to the slightly different samples.

\begin{deluxetable*}{lcD@{\(\pm\)}DD@{\(\pm\)}DD@{\(\pm\)}DD@{\(\pm\)}DD@{\(\pm\)}DD}
\tablefontsize{\footnotesize}
\tablecaption{Least Squares Analysis Results\label{tab:least_squares}}
\tablehead{
\colhead{Sample} &
\colhead{\(N\)} &
\multicolumn{4}{c}{\(U_\odot\)} &
\multicolumn{4}{c}{\(V_\odot\)} &
\multicolumn{4}{c}{\(W_\odot\)} &
\multicolumn{4}{c}{\(\gl_0\)} &
\multicolumn{4}{c}{\(\langle|z|^3\rangle/\langle|z|^2\rangle\)} &
\multicolumn{2}{c}{\(\sigma_V\)} \\
\colhead{} &
\colhead{} &
\multicolumn{4}{c}{\(\kms\)} &
\multicolumn{4}{c}{\(\kms\)} &
\multicolumn{4}{c}{\(\kms\)} &
\multicolumn{4}{c}{deg} &
\multicolumn{4}{c}{pc} &
\multicolumn{2}{c}{\(\kms\)}
}
\decimals
\startdata
\multicolumn{20}{l}{\citet{crovisier1978} Reproduction; \(|\gb| > 10^\circ\)} \\
\hline
\(|V_{\rm LSR}| < 40\kms\) &
294 &
-1.15 &
0.71 &
-0.20 &
0.83 &
-0.93 &
0.96 &
-10.95 &
2.96 &
132.74 &
14.07 &
6.81 \\
\(|V_{\rm LSR}| < 25\kms\) &
289 &
-1.49 &
0.65 &
0.29 &
0.77 &
-0.93 &
0.88 &
-12.01 &
3.04 &
119.78 &
12.96 &
6.23 \\
\(|V_{\rm LSR}| < 15\kms\) &
271 &
-1.18 &
0.53 &
0.69 &
0.63 &
-1.07 &
0.70 &
-15.09 &
3.57 &
85.56 &
10.73 &
4.87 \\
\hline
\multicolumn{20}{l}{\citet{crovisier1978} Reproduction; \(|V_{\rm LSR}| < 25\kms\)} \\
\hline
\(10^\circ < |\gb| < 20^\circ\) &
119 &
-0.33 &
1.07 &
0.70 &
1.21 &
-5.01 &
3.00 &
-11.24 &
4.06 &
111.72 &
16.45 &
7.18 \\
\(20^\circ < |\gb| < 40^\circ\) &
102 &
-2.16 &
0.95 &
-0.93 &
1.17 &
-1.37 &
1.30 &
-16.53 &
4.82 &
188.52 &
33.57 &
5.10 \\
\hline
\multicolumn{20}{l}{\citet{reid2019} Galactic Rotation Model; \(|\gb| > 10^\circ\)} \\
\hline
\(|V_{\rm LSR}| < 40\kms\) &
294 &
11.65 &
0.73 &
15.88 &
0.85 &
8.60 &
0.98 &
\multicolumn{4}{c}{\nodata} & 
128.04 &
14.88 &
6.97 \\
\(|V_{\rm LSR}| < 25\kms\) &
289 &
11.97 &
0.67 &
15.38 &
0.79 &
8.61 &
0.90 &
\multicolumn{4}{c}{\nodata} & 
114.25 &
13.82 &
6.40 \\
\(|V_{\rm LSR}| < 15\kms\) &
271 &
11.61 &
0.54 &
14.90 &
0.65 &
8.76 &
0.72 &
\multicolumn{4}{c}{\nodata} & 
77.85 &
11.62 &
5.03 \\
\hline
\multicolumn{20}{l}{\citet{reid2019} Galactic Rotation Model; \(|V_{\rm LSR}| < 25\kms\)} \\
\hline
\(10^\circ < |\gb| < 20^\circ\) &
119 &
10.77 &
1.11 &
15.11 &
1.23 &
12.75 &
3.09 &
\multicolumn{4}{c}{\nodata} & 
107.41 &
17.52 &
7.40 \\
\(20^\circ < |\gb| < 40^\circ\) &
102 &
13.15 &
0.99 &
16.28 &
1.24 &
8.45 &
1.37 &
\multicolumn{4}{c}{\nodata} & 
166.91 &
39.89 &
5.44 \\
\enddata
\end{deluxetable*}

\subsection{Updated Least Squares}

A modern re-analysis of the \citet{crovisier1978} data is warranted in
order to quantify the effect of an updated Galactic rotation
model. Here we adopt the A5 model of \citet{reid2019}, which is an
axisymmetric Galactic rotation curve fit to the positions and
kinematics of masers associated with high-mass star forming regions.
Since our sample of clouds is limited to the solar neighborhood, the
most important difference between this model and the local
approximation used in the previous analysis is the shape of the
rotation curve at the solar Galactocentric distance. This shape is
parameterized by Oort's \(A\) constant, which, for the
\citet{reid2019} model, is \(A \simeq 15.2\,\text{km s\(^{-1}\)
  kpc\(^{-1}\)}\) and not so different from the previous assumption.
Nonetheless, by incorporating this non-local Galactic rotation model,
we can be confident that second-order effects for potentially distant
\hi\ absorbing clouds are properly considered.

We use another least squares analysis to infer the model parameters by
minimizing the velocity residuals given by
equation~\ref{eq:least_squares} and again ignoring the LSR velocity
uncertainties of individual clouds. We fix \(R_0\) and the Galactic
rotation model parameters \(a_2\) and \(a_3\) to the mean point
estimates from the \citet{reid2019} A5 model but allow the parameters
defining the solar motion with respect to the LSR to vary. This least
squares method has one fewer free parameter (the nodal deviation) than
the original \citet{crovisier1978}
analysis. Table~\ref{tab:least_squares} shows the results of the least
squares analysis using the \citet{reid2019} Galactic rotation model
for the same subsets of data. Unsurprisingly, since the local shape of
the Galactic rotation curve is similar between both analyses, the
inferred distribution of \hi\ absorbing clouds is nearly the same
between the two methods.

\subsection{MCMC Moment Ratio}

There are two important deficiencies in the aforementioned least
squares analyses: outliers and uncertainties in the Galactic rotation
model. As noted by \citet{crovisier1978} and demonstrated in
Table~\ref{tab:least_squares}, the inferred shape of the
\hi\ absorbing cloud distribution strongly depends on the LSR
velocities included in the sample. This dependency suggests that the
data are not sufficiently described by the model; outliers bias the
inference. Furthermore, we use point estimates for the Galactic
rotation model parameters (i.e., Oort's \(A\) constant as in
\citet{crovisier1978}, or fixed \(R_0\), \(a_2\), and \(a_3\) in the
\citet{reid2019} A5 model), yet these parameters have associated
uncertainties that should be propagated into our inferred \(z\)
distribution. A Bayesian model is well-suited to overcome these
complications.

Here we develop a Bayesian model, parameterized in terms of the \(z\)
distribution moment ratio,
\(\langle|z|^3\rangle/\langle|z|^2\rangle\), to predict the LSR
velocity distribution of \hi\ absorbing clouds. As in the least
squares analyses, each cloud is assigned to its distance expectation
value following equation~\ref{eq:corrected}, at which we evaluate the
expected LSR velocity from the \citet{reid2019} A5 Galactic rotation
model, \(V_{\rm model}\). By assigning each cloud to its distance
expectation value, we are marginalizing over the latent distance of
each cloud, which we do not care to infer, under the assumption that
it is poorly constrained by the velocity. We assume that the random
component of the LSR velocity residuals is normally distributed with
variance \(\sigma_V^2\), so the likelihood that a cloud with observed
LSR velocity \(V_i\) is drawn from the model is a normal distribution
with mean \(V_i - V_{\rm model}\) and variance \(\sigma_V^2\). To
account for the possibility of outliers in our data, we include a
second, zero-centered, normally-distributed component in the
likelihood distribution with a variance \(\sigma_{V,\rm
  outlier}^2\). The total likelihood that cloud \(i\) is drawn from
this mixture model is thus
\begin{align}
  P(V_{i,\rm LSR} | \vect{w}, \vect{\theta}) = &\,w_{0}N(V_i - V_{\rm model}, \sigma_V^2 | \vect{\theta}) \nonumber \\
  & + w_{1}N(0, \sigma_{V,\rm outlier}^2)
\end{align}
where \(\vect{w} = (w_{0}, w_{1})\) are mixture weights that
quantify the fraction of data belonging to the model or outlier
population, respectively, such that \(w_{0} + w_{1} = 1\), and
\(\vect{\theta}\) is the set of model parameters that defines the
model, including \(\langle|z|^3\rangle/\langle|z|^2\rangle\) and the
Galactic rotation model parameters. We again ignore the LSR velocity
uncertainties of individual clouds in this likelihood calculation
since they are typically much smaller than \(\sigma_V\) and thus do
not contribute significantly to the likelihood determination.

In addition to the likelihood, we must also specify our prior
knowledge of the model parameters. For the \(z\) distribution moment
ratio, we adopt a \(k=2\) gamma distribution with a scale parameter
\(\theta = 50\,\text{pc}\). The gamma distribution is a convenient
specification both because it is broad (variance \(k\theta^2\)) -- and
thus encompasses a reasonable range of values -- as well as because it
approaches zero probability for unphysically small moment ratios.  For
the likelihood distribution widths, \(\sigma_V\) and \(\sigma_{V, \rm
  outlier}\), we use half-normal distributions with widths
\(10\,\text{km s\(^{-1}\)}\) and \(50\,\text{km s\(^{-1}\)}\),
respectively, which again are quite broad and encompass physically
reasonable expectations. The mixture weight prior is a Dirichlet
distribution with equal concentration. Finally, for the Galactic
rotation model parameters, we adopt as a prior a multivariate normal
distribution with means equal to the \citet{reid2019} A5 model
posterior point estimates and covariances inferred from the A5 model
posterior distribution standard deviations and correlation matrix
(Reid, M., private communication). Random samples from these prior
distributions (i.e., prior predictive checks) cover the range of
observed data, and thus we are confident that the chosen prior
distributions are reasonable.

We draw posterior samples from our Bayesian model using the
Hamiltonian Monte Carlo No-U-Turn sampler \citep[NUTS;][]{nuts2014}
implemented in \textit{pyMC}, a probabilistic programming framework in
\textit{python} \citep{pymc2023}. To diagnose convergence and ensure
that the sampler is not confined to local maxima of the posterior
distribution, we draw 500 tuning samples followed by 500 posterior
samples across four independent Markov chains. We check that the
chains are converged by computing the \(\hat{R}\) statistic, which
compares inter-chain variance to that across all chains
\citep{rhat2021}, as well as the effective sample size of each
parameter. In this and all subsequent MCMC analyses, we find \(\hat{R}
< 1.01\) and effective sample sizes \({>}1000\) for all parameters,
which is indicative of convergence. Furthermore, we draw random
samples from the posterior distribution (i.e., posterior predictive
checks) and confirm that the sampled posterior distribution is able to
reproduce the observations.

The results of our moment ratio MCMC analysis are listed in the first
column of Table~\ref{tab:mcmc}. Here, we use all of the
\hi\ absorption spectra with \(|b| > 10^\circ\); we do not exclude the
high-velocity features since our model is capable of identifying them
as outliers. The model parameters include the \(z\) distribution
moment ratio, \(\langle|z|^3\rangle/\langle|z|^2\rangle\), the
fraction of data belonging to the outlier distribution, \(f_{\rm
  outlier} \equiv w_1\), the width of the model and outlier
distributions, \(\sigma_V\) and \(\sigma_{V,\rm outlier}\),
respectively, and the parameters that describe the \citet{reid2019} A5
Galactic rotation model: the solar Galactocentric distance, \(R_0\),
the shape parameters, \(a_2\) and \(a_3\), and the solar motion with
respect to the LSR, \(U_\odot\), \(V_\odot\), and
\(W_\odot\). Although we do not sample the shape parameters directly,
we infer the Gaussian scale height, \(\sigma_z\), exponential scale
height, \(\lambda_z\), and rectangular distribution half-width,
\(W_{z, 1/2}\), from the moment ratio posterior distributions and
include them in Table~\ref{tab:mcmc}. A visual inspection of the
marginalized posterior distributions for each parameter reveals that
they are uni-modal and approximately normally distributed. The
parameter values and uncertainties in Table~\ref{tab:mcmc} represent
the posterior mean point estimates and 68\% highest posterior density
credible intervals, respectively.

\begin{deluxetable*}{lrlrlrlrl}
\tablefontsize{\footnotesize}
\tablecaption{MCMC Analysis Results\label{tab:mcmc}}
\tablehead{
\colhead{Model} &
\multicolumn{2}{c}{Moment Ratio} &
\multicolumn{2}{c}{Gaussian} &
\multicolumn{2}{c}{Exponential} &
\multicolumn{2}{c}{Rectangular}
}
\startdata
$N$ & \multicolumn{2}{c}{297}  & \multicolumn{2}{c}{297}  & \multicolumn{2}{c}{297}  & \multicolumn{2}{c}{297}  \\
$R_0$ (kpc) & $8.225$ & $^{+0.144}_{-0.134}$ & $8.220$ & $^{+0.146}_{-0.140}$ & $8.228$ & $^{+0.139}_{-0.138}$ & $8.222$ & $^{+0.158}_{-0.116}$ \\
$U_\odot$ (\kms) & $12.03$ & $^{+0.56}_{-0.48}$ & $12.02$ & $^{+0.55}_{-0.52}$ & $12.05$ & $^{+0.45}_{-0.63}$ & $12.00$ & $^{+0.51}_{-0.58}$ \\
$V_\odot$ (\kms) & $14.40$ & $^{+0.76}_{-0.66}$ & $14.36$ & $^{+0.76}_{-0.62}$ & $14.37$ & $^{+0.68}_{-0.82}$ & $14.37$ & $^{+0.63}_{-0.84}$ \\
$W_\odot$ (\kms) & $8.18$ & $^{+0.48}_{-0.56}$ & $8.19$ & $^{+0.47}_{-0.56}$ & $8.18$ & $^{+0.48}_{-0.59}$ & $8.20$ & $^{+0.53}_{-0.52}$ \\
$a_2$ & $0.955$ & $^{+0.048}_{-0.044}$ & $0.954$ & $^{+0.039}_{-0.053}$ & $0.956$ & $^{+0.052}_{-0.040}$ & $0.954$ & $^{+0.045}_{-0.043}$ \\
$a_3$ & $1.616$ & $^{+0.013}_{-0.014}$ & $1.616$ & $^{+0.015}_{-0.012}$ & $1.617$ & $^{+0.012}_{-0.015}$ & $1.617$ & $^{+0.012}_{-0.014}$ \\
$\sigma_V$ (\kms) & $5.22$ & $^{+0.45}_{-0.38}$ & $5.21$ & $^{+0.37}_{-0.46}$ & $5.23$ & $^{+0.38}_{-0.45}$ & $5.19$ & $^{+0.37}_{-0.44}$ \\
$f_{\rm outlier}$ & $0.133$ & $^{+0.034}_{-0.048}$ & $0.132$ & $^{+0.033}_{-0.051}$ & $0.130$ & $^{+0.032}_{-0.050}$ & $0.134$ & $^{+0.034}_{-0.047}$ \\
$\sigma_{V, \rm outlier}$ (\kms) & $22.9$ & $^{+2.3}_{-4.9}$ & $22.8$ & $^{+2.2}_{-4.9}$ & $23.0$ & $^{+2.2}_{-5.1}$ & $22.6$ & $^{+2.4}_{-4.8}$ \\
$\langle|z|^3\rangle/\langle|z|^2\rangle$ (pc) & $97.0$ & $^{+14.9}_{-15.0}$ & $98.4$ & $^{+15.4}_{-14.4}$ & $100.0$ & $^{+13.5}_{-15.8}$ & $95.7$ & $^{+12.4}_{-16.3}$ \\
$\sigma_z$ (pc) & $60.8$ & $^{+9.3}_{-9.4}$ & $61.7$ & $^{+9.6}_{-9.0}$ & \multicolumn{2}{c}{\nodata} & \multicolumn{2}{c}{\nodata} \\
$\lambda_z$ (pc) & $32.3$ & $^{+5.0}_{-5.0}$ & \multicolumn{2}{c}{\nodata} & $33.3$ & $^{+4.5}_{-5.3}$ & \multicolumn{2}{c}{\nodata} \\
$W_{z, 1/2}$ (pc) & $129.4$ & $^{+19.9}_{-20.0}$ & \multicolumn{2}{c}{\nodata} & \multicolumn{2}{c}{\nodata} & $127.5$ & $^{+16.6}_{-21.7}$ \\
log ELPD & \multicolumn{2}{c}{\nodata} & $-1017.5$ & $\pm20.3$ & $-1017.5$ & $\pm20.2$ & $-1017.5$ & $\pm20.3$ \\
\enddata
\end{deluxetable*}

\subsection{MCMC Shape Parameter}

If we assume the shape of the \(z\) distribution of \hi\ absorbing
clouds, then we can craft a Bayesian model to infer the shape
parameter of that distribution directly. Such a model is beneficial in
that it allows us to perform model comparison and determine which
distribution shape best represents the data. The only difference
between this model parameterization and the former is the relationship
between the \(z\) distribution shape parameter and the distribution
moment ratio. For a Gaussian, exponential, and rectangular
distribution parameterized by Gaussian scale height \(\sigma_z\),
exponential scale height \(\lambda_z\), and rectangular half-width
\(W_{z, 1/2}\), respectively, the \(z\) distribution moment ratio,
\(\langle|z|^3\rangle/\langle|z|^2\rangle\), is
\(2\sigma_z\sqrt{2/\pi}\), \(3\lambda_z\), and \(3W_{z, 1/2}/4\),
respectively. We use as a prior on the shape parameters a \(k=2\)
gamma distribution with a scale parameter \(\theta = 50\,\text{pc}\),
which again is a convenient distribution both because the prior
probability goes to zero for unphysically small shape parameters as
well as because the distribution is wide enough to cover reasonable
shape parameter values. Prior predictive checks demonstrate that our
choice of priors reasonably represents the data.

The results of our shape parameter MCMC analyses are listed in the
last three columns of Table~\ref{tab:mcmc} for the Gaussian,
exponential, and rectangular models. Again, we include all
\hi\ absorption features with \(|b| > 10^\circ\). We list the derived
\(z\) distribution moment ratio, although this parameter is not
sampled directly for these models. All models converge after 500
tuning samples and 500 posterior samples as determined by \(\hat{R} <
1.01\) for each parameter, an effective sample size \(>1000\) for each
parameter, and a visual inspection of the posterior predictive checks.
The marginalized posterior distributions are again uni-modal and
normally distributed in all cases, so the parameter values and
uncertainties in Table~\ref{tab:mcmc} represent the posterior mean
point estimates and 68\% highest posterior density credible intervals,
respectively.

Finally, we determine which of the three assumed \(z\) distribution
shapes best represents the data using leave-one-out cross-validation
\citep{vehtari2017} as implemented in \textit{ArviZ}, a
\textit{python} package for exploratory analysis of Bayesian models
\citep{arviz2019}. For each shape parameter model, the last row of
Table~\ref{tab:mcmc} lists the expected log pointwise predictive
density (ELPD), which is a measure of the out-of-sample predictive
fit. The model with the largest ELPD is preferred.

\section{Discussion}

\subsection{Vertical Distribution}

Although the statistical bias introduced by latitude truncation alters
the interpretation of the \citet{crovisier1978} analysis, the
statistical model that they develop to infer the vertical distribution
of \hi\ absorbing clouds is robust. Based on our least squares
implementation of this model and using the original \citet{nancay1978}
data, we are able to reproduce the \citet{crovisier1978} results and
infer the shape of the \(z\) distribution of cold
\hi\ clouds. Depending on the LSR velocity selection of the
sub-sample, we infer a \(z\) distribution shape defined by the ratio
of the third moment (skewness) to the second moment (variance),
\(\langle|z|^3\rangle/\langle|z|^2\rangle \simeq 110 \pm
15\,\text{pc}\) (see Table~\ref{tab:least_squares}). For an assumed
Gaussian distribution, this moment ratio corresponds to a scale height
\(\sigma_z \simeq 70\,\text{pc}\), which is a factor of two smaller
than that inferred by \citet{crovisier1978} due to the latitude
truncation bias discussed in Section~\ref{sec:mistake}.

\begin{figure}
  \centering
  \includegraphics[width=\linewidth]{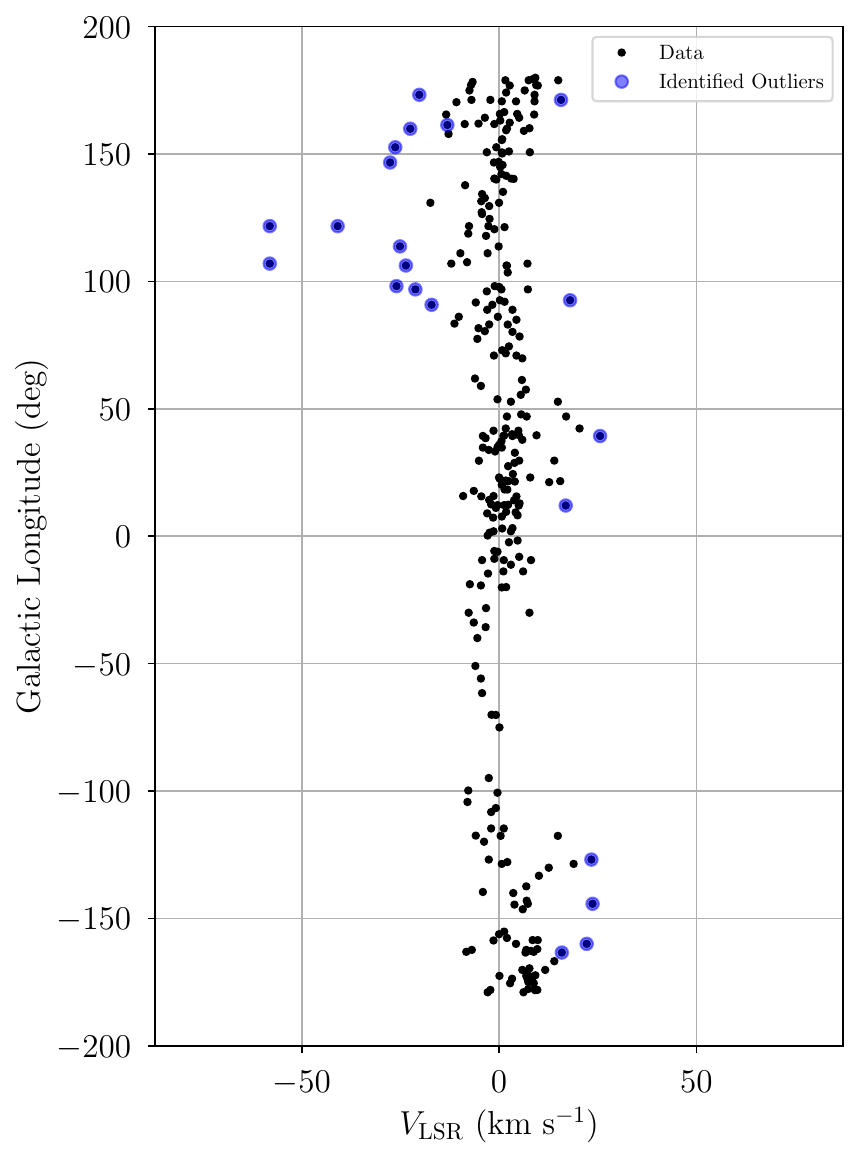}
  \caption{Galactic longitudes and LSR velocities, \(V_{\rm LSR}\), of
    the \citet{nancay1978} \hi\ absorption detections. Those data with
    a \(>50\%\) posterior probability of being outliers in the moment
    ratio model are highlighted in blue. \label{fig:outliers}}
\end{figure}

The least squares analysis results are sensitive to the LSR velocity
criteria of each subsample, which implies that the data include
outliers that are not explained by the model. Therefore, we develop
two Bayesian models -- one to infer the \(z\) distribution moment
ratio and one to infer the shape parameter of an assumed distribution
directly -- that include an outlier component in the likelihood
distribution. We find that each model converges to a similar \(z\)
distribution moment ratio, \(\langle|z|^3\rangle/\langle|z|^2\rangle
\simeq 97\pm 15\,\text{pc}\), which corresponds to a Gaussian scale
height \(\sigma_z = 61\pm 9\,\text{pc}\), an exponential scale height
\(\lambda_z = 32\pm 5\,\text{pc}\), and a rectangular half-width
\(W_{z, 1/2} = 129\pm 20\,\text{pc}\) (see Table~\ref{tab:mcmc}).  The
ELPD is nearly identical between models, which indicates that the data
are not able to distinguish between these three assumed distributions.
Furthermore, we find that \({\sim}13\%\) of the \citet{nancay1978}
\hi\ absorption detections appear to be poorly explained by our model
and instead belong to an outlier population. Figure~\ref{fig:outliers}
shows the Galactic longitude-velocity distribution of the
\citet{nancay1978} \hi\ absorption detections, where we highlight
those data with a \(>50\%\) probability of being a member of the
outlier population based on samples from the posterior distribution of
the moment ratio model. Clearly, the outliers belong to a population
of intermediate-velocity \hi\ absorption features. These detections
may represent a collection of intermediate-velocity clouds
\citep[e.g.,][]{wakker2004,lehner2022}, perhaps the remnants of
infalling high-velocity clouds \citep[e.g.,][]{begum2010}, but we
defer further investigation to a future work.

\subsection{Implications}

The fact that we have reduced the apparent vertical distribution of
cold \hi\ clouds by a factor of two compared to the original
\citet{crovisier1978} analysis has potentially significant
implications for our understanding of hydrostatic balance in the solar
neighborhood ISM. Our result suggests that, in the solar neighborhood,
the scale height of the CNM is nearly the same as that of the
molecular gas layer \citep[\(\sigma_z \simeq 50\,\text{pc}\),
  e.g.,][]{heyer2015}. This complicates the already-limited
characterization of the vertical distribution of the CNM given the few
observational constraints \citep{mcclure-griffiths2023} and draws into
question the claimed variation in CNM scale height with Galactocentric
distance.

It is possible that the solar neighborhood gives a biased view of the
vertical distribution of CNM clouds. Consider, for example, the
analysis of \citet{dickey2022}, which uses \hi\ absorption detections
near \(0\kms\) to infer the vertical distribution of cold \hi\ clouds
near the solar Galactocentric distance. They find a latitude
distribution of \hi\ absorption features composed of two Gaussian
distributions, the broader of which they attribute to local \hi\ and
the narrower of which they attribute to \hi\ at the solar
Galactocentric distance on the far side of the Galaxy. Given a
heliocentric distance \(d \simeq 15.6\,\text{kpc}\), the latitude
distribution width of \({\sim}0^{\circ}.59\) corresponds to a \(z\)
distribution scale height \(\sigma_z = 160\,\text{pc}\). This is more
than a factor of two greater than our inferred CNM scale height in the
solar neighborhood. Since \citet{dickey2022} select their sample
kinematically, they may be biased by intermediate-velocity cloud
interlopers that exist at a variety of distances and thus complicate
the interpretation of the latitude distribution. It is also possible,
however, that the CNM scale height varies due to local effects like
the stellar surface density and star formation history.

Recent isolated galaxy simulations have attempted to overcome the
resolution and physical limitations of galaxy evolution models in
order to predict realistic distributions of different ISM phases on
sub-parsec scales. One such simulation, that of \citet{smith2023},
uses the hydrodynamic galaxy models of \citet{tress2020} and
\citet{tress2021} to reach parsec resolution (or better) in the cold
gas. The \citet{smith2023} simulation assumes a constant metallicity
across the disk, but they test the effect of a varying FUV
interstellar radiation field that is tied to the time-averaged star
formation rate surface density distribution compared to a constant
radiation field. In both cases, they find that the CNM scale height is
smaller than the total \hi\ scale height everywhere outside of the
Galactic center. This is in contrast to the \citet{dickey2009} study
which finds a similar scale height between the WNM and CNM in the
outer Galaxy. Furthermore, the \citet{smith2023} simulations
demonstrate that, although the azimuthally-averaged total \hi\ scale
height increases with Galactocentric distance, the
azimuthally-averaged CNM scale height remains roughly constant at
\(\sigma_z \simeq 50\,\text{pc}\) for the varying radiation field and
increases slightly from \({\sim}50\,\text{pc}\) to
\({\sim}100\,\text{pc}\) for the constant radiation field. These
simulations are thus consistent with our solar neighborhood inference,
although their Figure 10 suggests that there could be local variations
in the CNM scale height on the order of a factor of
\({\sim}2\). Such local variations could explain the
\citet{dickey2022} scale height at the solar Galactocentric distance
in the fourth quadrant.

\subsection{Galactic Rotation}

In addition to the vertical distribution of \hi\ absorbing clouds, our
Bayesian model also infers the \citet{reid2019} Galactic rotation
model parameters, including the solar Galactocentric distance. We use
as priors a multivariate normal distribution defind by the means,
standard deviations, and correlation coefficients of the
\citet{reid2019} A5 model posterior distribution. Except for the solar
peculiar velocity components, our model posterior distribution means
match the prior distribution means within the uncertainties and the
widths of the posterior distributions are similar to those of the
priors (see Table~\ref{tab:mcmc}). These two facts suggest that the
\citet{nancay1978} data are not able to further constrain \(R_0\) and
the Galactic rotation model parameters. On the other hand, both our
least squares analysis and Bayesian models converge to similar
estimates for the solar peculiar velocity components, \((U_\odot,
V_\odot, W_\odot) = (12.0\pm0.5, 14.4\pm0.7, 8.2\pm0.5)\), which are
more precise than the prior distributions. These parameter estimates
are consistent with those inferred from stellar kinematics
\citep{schonrich2010,zbinden2019}, although the stellar results appear
sensitive to both the method of analysis and the stellar sample
\citep[e.g.,][]{ding2019}.

\subsection{Code Package}

The Bayesian model, least squares, and MCMC methods used this work are
provided to the community as an open source package:
\texttt{kinematic\_scaleheight}\footnote{\url{https://doi.org/10.5281/zenodo.10818724}}
\citep{kinematic_scaleheight}. Although we have only demonstrated it
as applied to \hi\ absorption data, it is possible to use these models
and algorithms to infer the vertical distribution of any tracer in the
solar neighborhood for which there is a reasonable expectation that
the tracer follows a given Galactic rotation model.

\section{Conclusion \& Future Work}

The vertical distribution of the ISM, perpendicular to the Galactic
plane, is a constraint on models of ISM hydrostatic balance and galaxy
evolution. In particular, the distribution of CNM clouds appears to be
set gravitationally by the stellar surface density and kinematically
by the energy injection from star formation
\citep{mckee1977,smith2023}.  Due to the observational challenges
associated with identifying individual CNM clouds, few studies have
explored their vertical distribution across the Galactic disk. In the
solar neighborhood, the statistical analysis of \citet{crovisier1978}
and the extension by \citet{belfort1984} are the only such studies.

We identify a bias in the \citet{crovisier1978} method that results in
an overestimate of the scale height of local CNM clouds by a factor of
\({\sim}1.5\) to \({\sim}3\). With a corrected least squares method as
well as a novel Bayesian model, we use the \citet{nancay1978}
\hi\ absorption data to infer that the vertical distribution of
\hi\ absorption in the solar neighborhood is best described by
\(\langle|z|^3\rangle/\langle|z|^2\rangle = 97\pm 15\,\text{pc}\),
where \(\langle|z|^3\rangle\) and \(\langle|z|^2\rangle\) are the
third (skewness) and second (variance) moments of the distribution,
respectively. This moment ratio corresponds to a Gaussian scale height
\(\sigma_z = 61\pm 9\,\text{pc}\), an exponential scale height
\(\lambda_z = 32\pm 5\,\text{pc}\), and a rectangular half-width
\(W_{z, 1/2} = 129\pm 20\,\text{pc}\), although the \citet{nancay1978}
data are unable to distinguish between these three possible vertical
distribution shapes.  Furthermore, our model suggests that
\({\sim}13\%\) of the local \hi\ absorption detections may belong to a
population of intermediate-velocity cloud interlopers, since they are
not well-described by our model. Both the least squares and Bayesian
models converge to estimates of the solar peculiar motion that are
consistent with those inferred from stellar kinematics
\citep[e.g.,][]{schonrich2010}: \((U_\odot, V_\odot, W_\odot) =
(12.0\pm0.5, 14.4\pm0.7, 8.2\pm0.5)\). Our analysis tools are provided
to the community as an open-source software package
\citep{kinematic_scaleheight}.

The scale height of CNM clouds in the solar neighborhood is similar to
both the inferred inner-Galaxy CNM scale height \citep[\(\sigma_z
  \simeq 50-90\,\text{pc}\), e.g.,][]{dickey2022} as well as the scale
height of the molecular gas layer in the solar neighborhood
\citep[\(\sigma_z \simeq 50\,\text{pc}\), e.g.,][]{heyer2015}. These
results are consistent with simulations, which suggest that the scale
height of cold \hi\ should remain roughly constant out to the edge of
the star forming disk \citep{smith2023}. At the solar Galactocentric
distance on the far side of the Galaxy, however, \citet{dickey2022}
find a \(2-3\times\) larger scale height. Either there are significant
variations in the scale height due to local effects, or their
inference is biased due to intermediate-velocity \hi\ absorption
interlopers that complicate the interpretation of their
kinematically-selected sample. A comprehensive analysis of modern
\hi\ absorption observations in the solar neighborhood is warranted
and will be the subject of an upcoming study.

\section*{Acknowledgments}

This research has made use of NASA's Astrophysics Data System
Bibliographic Services. T.V.W. is supported by a National Science
Foundation Astronomy and Astrophysics Postdoctoral Fellowship under
award AST-2202340. D.R.R. is supported by a National Science
Foundation Astronomy and Astrophysics Postdoctoral Fellowship under
award AST-2303902. The authors acknowledge Interstellar Institute's
program ``II6'' and the Paris-Saclay University's Institute Pascal for
hosting discussions that nourished the development of the ideas behind
this work, as well as J. M. Dickey for providing comments that
improved the quality of this paper.

\software{
  ArviZ \citep{arviz2019},
  pyMC \citep{pymc2023},
  Matplotlib \citep{matplotlib2007},
  NumPy \& SciPy \citep{numpyscipy2011},
  \texttt{kinematic\_scaleheight} \citep{kinematic_scaleheight},
}

\bibliography{bibliography}

\end{document}